# Quantum oscillations and linear magnetoresistance in ultraclean CaVO$_3$ thin films

Mahni Müller[1], Maria Espinosa[1], Olivio Chiatti[1], Tatiana Kuznetsova[2], Roman Engel-Herbert[2,3] and Saskia F. Fischer[1,4]

[1] Novel Materials Group, Institute of Physics, Humboldt-Universität zu Berlin, Newtonstr. 15, 12489 Berlin, Germany

[2] Department of Materials Science and Engineering, The Pennsylvania State University, University Park, PA 16802 USA

[3] Paul Drude Institute for Solid State Electronics, Hausvogteiplatz 5-7, 10117 Berlin, Germany.

[4] Center of the Sciences of Materials Berlin, Humboldt-Universität zu Berlin, Zum Großen Windkanal 2, 12489 Berlin, Germany

Advances in epitaxy of transition metal oxides with perovskite structure allow novel insights into transport mechanisms of strongly correlated electron systems, which are of interest for future transparent electronics. In this study, we investigate magnetotransport properties of thin epitaxial CaVO$_3$ films, grown coherently strained on LaAlO$_3$, and demonstrate for the ultraclean limit quantum oscillations. Fermi liquid behavior is detected in the temperature-dependent resistivity $\rho(T) \sim T^2$ up to 20 K, with effective mean free paths exceeding up to 20 times the film thickness (38 nm). Shubnikov-de Haas oscillations and a non-linear Hall resistance reveal two electron-like (1 and 2) and one hole-like (h) carriers, reflecting the three-fold Fermi surface of orthorhombic CaVO$_3$, with effective charge carrier densities and mobilities at 4.2 K of: (1) $n_1 \approx 9.3 \cdot 10^{21}$ cm$^{-3}$ with low mobility $\mu_1 \approx 926$ cm$^2$ V$^{-1}$s$^{-1}$, (2) $n_2 \approx 7.2 \cdot 10^{19}$ cm$^{-3}$ with high mobility $\mu_2 \approx 6600$ cm$^2$ V$^{-1}$s$^{-1}$, and (h) $n_h \approx 2.2 \cdot 10^{18}$ cm$^{-3}$ with $\mu_h \approx 1500$ cm$^2$ V$^{-1}$s$^{-1}$. A non-saturating linear magnetoresistance dominates at low temperatures, exceeding the value for single crystals by 30%. Our findings on epitaxial films demonstrate the delicate interplay of multiple carriers with correlations stemming from a non-spherical nested Fermi surface of a perovskite structure with orthorhombic distortion.

## I. INTRODUCTION

Transparent conductors are in high demand for optoelectronics and require both high transparency and high conductivity[1-6]. Typical doped wide band-gap semiconductors (e. g. ITO, ZnO) face limitations due to the increased charge carrier scattering[7] and are aimed to be replaced by more abundant materials [8]. Correlated metals offer high electrical conductivity and optical transparency, due to increased electron-electron correlations and a high effective mass[9]. $CaVO_3$ and $SrVO_3$ are correlated metals with an $ABO_3$ perovskite-type crystal structure and a $3d^1$ electron configuration, and are often described as Fermi liquids[9,10]. While the crystal structure of $CaVO_3$ is orthorhombic, $SrVO_3$ is cubic. Despite its smaller lattice parameter and thus larger orbital overlap, the orthorhombic distortion is expected to decrease the electronic bandwidth, which increases electron localization[11–13] and hence the effective mass[14], placing $CaVO_3$ closer to the Mott transition[2,11]. Improvements of oxide thin film growth allows synthesizing epitaxial films of highest quality with low defect concentration[15]. Recent reports[16] on the intrinsic transport properties of $SrVO_3$ in the ultraclean limit reveal deviations from the normal Landau-Fermi liquid description, by detection of an anomalously high Hall coefficient and a scattering time anisotropy at 4 K. Studies using photoemission spectra (PES) and local density approximation and dynamical mean-field theory (LDA + DMFT), have revealed that the electronic structure and Fermi surface geometry of $CaVO_3$ and $SrVO_3$ are similar[19,20]. Calculations reveal a three-fold Fermi surface[9,17,18]. To date, several open issues persist: for thin films no quantum oscillations[2,27] have been reported; the differences between $CaVO_3$ and SrVO$_3$ in correlation effects and the origin of a non-saturating linear magnetoresistance in both call for further investigations.

In this study, we demonstrate that quantum oscillations, which are intrinsic to CaVO$_3$ single crystals[2,27], are also observed in high-quality epitaxial thin films and, further, that an enhanced non-saturating linear magnetoresistance appears at low temperatures. Systematically, we investigate the temperature-dependent magnetotransport of a series of CaVO$_3$ thin films with increasing crystal quality, quantified by the residual resistivity ratio *RRR* (from 2 to 90). From the temperature-dependent resistance the scattering mechanisms are analyzed via the exponent $\alpha$ of a power-law behavior and the Fermi liquid behavior is discussed for $\alpha = 2$. An approximative two carrier fit to the magnetoresistance and Hall resistance is performed at lower magnetic fields up to 6 T. The full three Fermi surface theory is approached by the identification of a third carrier channel at higher fields up to 12 T. We discuss the outcomes with previous results published for CaVO$_3$ single crystals[1,2,27] and comparable SrVO$_3$ thin epitaxial films[16], showing that the transport characteristics from CaVO$_3$ epitaxial films of highest quality differ from both. The results indicate how transport properties can be tailored by size control of coherently applied small strain and/or increased surface scattering in epitaxial thin films of these correlated metals.

## II. RESULTS

The structural properties of the films under various growth conditions were investigated in detail in Ref. 21. Growth control and surface reconstruction was monitored by RHEED patterns for determination of film surface crystallinity and smoothness. Atomically flat films with terrace-like morphology and rms roughness values as low as 1.32 Å were found for films grown at VTIP pressures between 75 and 83 mTorr. This is the growth window for the epitaxial films of the highest quality, as confirmed by transport measurements in Ref. 21 and this work.

Here, we present the temperature-dependent resistivity of CaVO$_3$ films synthesized under these different growth conditions in Fig. 1(a). The residual resistivity ratio $RRR = \frac{\rho(300 \text{ K})}{\rho(4.2 \text{ K})}$ was determined, which is a measure for the crystalline quality in macroscopic samples and serves as a lower limit for size-constrained mesoscopic transport in thin film layers[29]. Considering the film thickness of 38 nm, the maximum value of $RRR = 98$ for CaVO$_3$ on LaAlO$_3$ is comparable to that of bulk single crystals[1,2] ($RRR_{\text{max}} \approx 200$) and thicker films[26,28] of CaVO$_3$ on SrTiO$_3$ ($RRR_{\text{max}} \approx 1200$). In the ultraclean limit, we determine a minimal residual resistivity of $\rho(T = 4.2 \text{ K}) \approx 3.6 \cdot 10^{-7} \Omega$ cm ($RRR = 98$).

An approximate expression of the temperature-dependent resistivity $\rho(T) = \rho_0 + A \cdot T^\alpha$ is used in order to analyze the scattering mechanisms, with the exponent $\alpha(T) = \frac{\partial \log(\rho(T) - \rho_0)}{\partial \log T}$, as shown in Fig. 1(b). At room temperature $\alpha \approx 2$ is found for all films. With decreasing temperature, $\alpha$ increases and reaches a maximum of $\alpha = 2.8$ to $3.2$ around 30 K and 60 K for films in the ultraclean and disordered limit, respectively. Below 20 K, the films with $RRR > 90$ show an $\alpha \approx 2$ dependence, whereas the films with $RRR \leq 10$ show a varying $\alpha < 3$. Below $T \approx 1$ K all films reach a constant residual resistivity $\rho_0$ (see also Supplementary Material S1). As depicted in Fig. 1(b), $\alpha$ for a SrVO$_3$ thin film in the ultraclean regime[16] agrees well with the temperature behavior for equivalent CaVO$_3$ films ($RRR > 68$).

A change in the exponent $\alpha$ indicates a crossover of the dominant scattering mechanism, from which we identify three temperature regimes: (i) At low temperatures ($T < 20$ K) electron-electron scattering dominates and results in a $T^2$-dependence[33]. This has been observed for CaVO$_3$ single crystals[1,2] and thin films[24,26,28], and for epitaxial SrVO$_3$ thin films [16,34]. However, for lower crystalline quality ($RRR = 2$) our results indicate that the low-temperature resistivity is dominated by impurity scattering. (ii) At intermediate temperatures the temperature dependence of $\alpha$ is similar to that of transition metals with to *s-d* scattering, often associated with the Bloch-Wilson limit[40-42] ($\alpha = 3$). For instance, high-purity vanadium samples[41] exhibit a peak in $\alpha \approx 3$ at 40 K (see Supplementary Material S2). In CaVO$_3$, the density of states of the V 3$d$ and O 2$p$ orbitals near the Fermi energy[19,31] suggests that a similar physical mechanism could result in a $T^3$-dependence. (iii) At higher temperatures $\alpha \approx 2$ is observed above 200 K (Fig. 1(b)) and for SrVO$_3$ thin epitaxial films[34,38] above 100 K. Theoretically, an increased Hubbard-Coulomb interaction increases the effective electron-phonon coupling[36,37], increasing the electron-phonon scattering and the effective mass[34] $m^*_{\text{e-ph}}$. Further, dynamical mean field theory calculations[38] for SrVO$_3$ attribute $\alpha \approx 2$ at higher temperatures to electron-phonon scattering with an increased contribution from optical phonons[38]. Other proposals[34] discuss $\alpha \approx 2$ at higher temperatures in terms of electron-phonon scattering on cylindrical Fermi surfaces[39] or of polaronic transport due to an enhanced electron-phonon coupling[34]. We propose that our observations for CaVO$_3$ epitaxial thin films share common physical origins with those in SrVO$_3$, given the similarities in band structure, Fermi surfaces[19,20] and Debye temperature $\theta_D$, despite discrepancies in absolute values reported in the literature[3,31] for $\theta_D$.

The temperature dependence of transport material parameters is depicted in Fig. 1(c) and Fig. 1(d). For very low magnetic fields ($|B| \leq 700$ mT) the Hall resistance $R_{xy}$ was measured and the Hall coefficient was determined as $R_H = \frac{\partial R_{xy}}{\partial B}$. A first characterization is obtained by analyzing these low-field magnetotransport data within a one-band model approximation: the *effective* carrier density $n_{\text{eff}}$ is the inverse Hall coefficient, $n_{\text{eff}} = (-eR_H)^{-1}$, and the *effective* mobility $\mu_{\text{eff}}$ is given by $\mu_{\text{eff}} = (e\, n_{\text{eff}}\, \rho)^{-1}$, where $\rho$ is the resistivity in zero magnetic field. When evaluated in this manner, below 100 K the CaVO$_3$ films show a strong increase in $\mu_{\text{eff}}(T)$ with decreasing temperature, consistent with a decrease in electron-phonon scattering. For the highest *RRR* the largest increase in $\mu_{\text{eff}}$ is observed. While the effective carrier density $n_{\text{eff}}$ at room temperature is nearly the same for all films, its significant temperature-dependent decrease is strongly *RRR* dependent. For highest *RRR* the largest decrease in $n_{\text{eff}}$ of 33 % from room temperature to 4.2 K is observed. From the values of $\mu_{\text{eff}}(T)$ and $n_{\text{eff}}(T)$ a temperature-dependent *effective* electronic mean free path (MFP) $l_{\text{e}} = \frac{\hbar}{e}\mu_{\text{eff}} \cdot (3\pi^2 n_{\text{eff}})^{\frac{1}{3}}$ can be derived, as shown in Fig. 1(d). For high $RRR$ a value of up to $l_{\text{e}} \approx 860$ nm at 4.2 K is determined, exceeding the film thickness of $d = 38$ nm by an order of magnitude. Therefore, elastic scattering at the surfaces and interfaces of the films plays an important role for $RRR > 26$, as compared to CaVO3 single crystals. This leads to anisotropic scattering by extrinsic size effects. However, intrinsic anisotropies due to the band structure require more than a one-band model, which does not account for the three-fold Fermi surface in CaVO3 and its anisotropies. Intrinsically, scattering times for different bands may vary significantly over the Fermi surface, as has been discussed for SrVO3[16].

Magnetotransport at higher fields up to 6 T provides more detailed information for a multi-band model. The longitudinal magnetoresistance $MR = \frac{R_{xx}(B)}{R_{xx}(B=0)} - 1$ is shown in Fig. 2, for higher epitaxial quality ($RRR = 60$) in Fig. 2(a) and for lower quality ($RRR = 2$) in Fig. 2(e). A striking change in the $MR$ for $RRR = 60$ is observed in magnitude when cooling from 80 K down to 4 K (with a maximum of 4 % at 80 K and of 44% at 4 K), which can be related to the increase of $\mu_{\text{eff}}$ with decreasing temperature. Accordingly, the field dependence of the magnetoresistance displays increasingly linear behavior below $T < 20$ K. The data were fitted with $MR = a_{MR}|B| + b_{MR} \cdot B^2 + c_{MR} \cdot |B^3|$ with linear, quadratic and cubic $MR$ coefficients $a_{MR}, b_{MR}$ and $c_{MR}$, shown in Fig. 2(b)-(d), respectively. The cubic term improves the fit and may arise from effects at higher fields, as discussed below. The linear coefficient $a_{MR}$ is negligible at higher temperatures above 60 K and increases continuously with decreasing temperature. In contrast, the quadratic coefficient $b_{MR}$ has a maximum around $T \approx 30 - 40$ K, which coincides with the maximum exponent $\alpha$ of the temperature-dependent resistivity. At temperatures below 40 K the linear component dominates and we find $a_{MR} \cdot |B| \gg b_{MR} \cdot B^2 \gg |c_{MR} \cdot B^3|$.

The corresponding $MR$ data for the lowest epitaxial quality ($RRR = 2$) is shown in Fig. 2(e). In the field dependence, the $MR$ does not display a dominant linear behavior below $T < 20$ K. The fitting coefficients are shown in Fig. 2(f)-(h) and reveal a qualitatively similar temperature dependence as for the films of higher epitaxial quality, but with magnitudes of a factor 100 smaller for $a_{MR}$ and a factor 10 smaller for $b_{MR}$.

These features of the field dependence can be clearly observed when $MR$ is plotted as a function of $\mu_{\text{eff}}B$, as shown in Fig. 3(a) and b. For $RRR = 60$ (Fig. 3(a)) three temperature regimes become obvious: (I) the low-temperature regime with a dominant linear $MR$ below 20 K; (II) the intermediate temperature regime, displaying a significant linear contribution to $MR$ (from 20 K to 60 K); and (III) the high-temperature regime, with a dominant quadratic term above 60 K. Instead for the disordered limit $RRR = 2$ (Fig. 3(b)), the field dependences of all temperatures resemble regime (III) of Fig. 2(a) ($\beta \approx 1.58$). This indicates that intrinsic multi-band contributions become evident for higher epitaxial quality at lower temperatures, and interestingly the temperature regimes below 20 K (i) from the zero-field resistivity and (I) from the magnetoresistance coincide.

Further, a double-logarithmic plot of $MR$ as a function of $\mu_{\text{eff}}B$, as depicted in Fig. 3(c), clearly shows the power-law behavior $MR \sim B^{\beta}$ and a crossover of the exponents from the temperature regime (I): $\beta \approx 1$ to (III): $\beta \approx 1.65$. This is detailed in Fig. 3(d) for both $RRR = 60$ and $RRR = 2$, showing further the striking difference of a factor 100 of the $MR$ at low temperatures between high and low epitaxial quality films.

The temperature-dependent transversal Hall resistance $R_{xy}$ up to 6 T for the higher quality epitaxial film ($RRR = 60$) shows non-linearities below 100 K, indicating multi-carrier transport. In Fig. 4(a), these are displayed by the numerical derivative of the Hall resistance $\frac{\partial R_{xy}}{\partial B}$. It is nearly constant from room temperature down to 100 K. In the range 100 K $> T >$ 60 K, a small maximum can be seen at $B = 0$, consistent with a small hole contribution. Below 60 K a minimum appears at zero magnetic field, which increases in magnitude down to 4.2 K. This is indicative of the appearance of a second electron channel. In contrast, the film with $RRR = 2$ shows a constant derivative at all temperatures (see Supplementary Material S3), which indicates that defect-induced scattering limits the mobilities and mask multi-carrier transport phenomena.

Because the Fermi surface in CaVO$_3$ is triply degenerate, we attribute these changes in the Hall slope to carriers from different bands[31]. From Fig. 4(a) the contribution of a hole channel is expected to be negligible, therefore we apply first a two-band model with $R_{xy}(B) = -\frac{B}{e} \cdot \frac{(n_1\mu_1^2 + n_2\mu_2^2) + \mu_1^2\mu_2^2(n_1+n_2)B^2}{(n_1\mu_1 + n_2\mu_2)^2 + \mu_1^2\mu_2^2(n_1+n_2)^2B^2}$ (see Supplementary Material S4 and S5). From the fits we obtain electron mobilities and densities as shown in Fig. 4(b)-(c). The results indicate one channel with a large electron density $n_1$ and low mobility $\mu_1$, and a second channel with a smaller electron density $n_2$ and higher mobility $\mu_2$. At $T = 4.2$ K we obtain: for channel 1, $n_1 \approx 9.3 \cdot 10^{21}$ cm$^{-3}$ and $\mu_1 \approx 926$ cm$^2$ V$^{-1}$ s$^{-1}$; for channel 2, $n_2 \approx 7.2 \cdot 10^{19}$ cm$^{-3}$ and $\mu_2 \approx 6600$ cm$^2$ V$^{-1}$ s$^{-1}$. The temperature dependence of $n_1$ is similar to the effective carrier density $n_{\text{eff}}$, indicating that channel 1 is the dominant transport contribution. The electron density of channel 2 is $n_2 \approx 1.5 \cdot 10^{18}$ cm$^{-3}$ for temperatures above $T \approx 50$ K and increases with decreasing temperature. The linear contribution to the magnetoresistance $a_{MR}$ and the electron density of channel 2 have similar temperature dependences, therefore indicating that channel 2 is the dominant contribution to the linear magnetoresistance and responsible for the minimum at $B \approx 0$ in the Hall slope. Both channels show similar temperature dependence of the mobility. Therefore, the two-band model appears sufficient to describe the non-linear Hall effect at intermediate magnetic fields in CaVO$_3$ high-quality epitaxial layers. While non-linearities in the Hall resistance have been previously found in SrVO$_3$ epitaxial layers in the ultraclean limit, those results differ because a pronounced hole contribution was observed down to low temperatures [16].

In order to test whether the magnetoresistance saturates and quantum oscillations for $\mu_{\text{eff}}B \gg 1$ can be observed, high-field transport measurements up to 12 T were performed. In Fig. 5(a) and (b) the magnetoresistance and Hall resistance data, respectively, are shown for the CaVO$_3$ thin film with $RRR = 64$ in the temperature range from 100 K to 50 mK. Strikingly, the magnetoresistance does not saturate up to 12 T. A non-saturating linear magnetoresistance (LMR) has been reported for single crystals[1], where magnetoresistance and magnetization anisotropies have been accounted for by an open orbit along the (110) direction, leading to an absence of saturation of the positive magnetoresistance.

In Fig. 5(c) the temperature dependence of the numerical derivative of the magnetoresistance reveals several features: at 100 K the quadratic field dependence is dominant, while at the lowest temperature of 50 mK the low-field regime ($< 0.2$ T) is predominantly quadratic and therefore distinct from the high-field regime ($> 0.2$ T) of a dominant linear component with a minor quadratic contribution. Below 45 K the $MR$ slope shows the onset of quantum oscillations at fields above 7 T. In Fig. 5(d), the temperature dependence of the numerical derivative of the Hall resistance confirms this, such as the maximum at $B = 0$ for $T > 60$ K, the minimum for $T < 60$ K and the quantum oscillations above 7 T and below 45 K. Two-band fits yield results consistent with those in Fig. 4.

The quantum oscillations in the $MR$ and Hall slopes at 50 mK are field-symmetric, as depicted in Fig. 6(a) (as a function of the absolute value of the magnetic field $|B|$). The Hall slope oscillation is periodic in $\frac{1}{B}$, as shown in Fig. 6(b), with a period of $\Delta\left(\frac{1}{B}\right) = \frac{0,019}{\text{T}}$. This is consistent with Shubnikov-de-Haas (SdH) oscillations of a frequency $B_{\text{SdH}} \approx 52$ T. From $\Delta\left(\frac{1}{B}\right) = \frac{2\pi e}{\hbar} S_F^{-1}$ we determine a carrier density [32], assuming that $S_F = \pi k_F^2 = \pi(3\pi^2 n_{SdH})^{\frac{2}{3}}$ is the relevant cross-section of the Fermi surface. The corresponding carrier density is $n_{SdH} \approx 2.2 \cdot 10^{18}$ cm$^{-3}$, much smaller than the electron densities from the two-band fits. From the results of Fig. 4 and 5, a hole channel with smaller carrier density than both electron channels is expected. Therefore, the hole channel may account for the observed quantum oscillations. The corresponding mobility can be estimated from the lowest magnetic field $B_{min}$, where quantum oscillations are observed: $\mu_h = (B_{min})^{-1} \approx 1500$ cm$^2$/Vs. These results confirm the excellent quality of the epitaxial CaVO$_3$ layers in the ultraclean limit. SdH oscillations are known from CaVO$_{3-y}$ single crystals[2] ($RRR = 20, B_{SdH} = 300$ T), however, for ultraclean SrVO$_3$ epitaxial layers ($RRR = 195$) no SdH oscillations were reported[16] up to 18 T.

### III. DISCUSSION

In the following we discuss our experimental results in the *ultraclean* limit in the context of previous reports. In general, the temperature dependence of the resistivity $\rho$ of our epitaxial 38 nm thin films of CaVO$_3$ on LaAlO$_3$ resembles that of bulk CaVO$_3$ single crystals[1,2] and is similar to that of ultraclean SrVO$_3$ epitaxial layers[16]. We identify three temperature regimes with power law behavior (exponent $\alpha$) of the correlated metal: (i) Fermi liquid behavior ($\alpha \approx 2$) at low temperatures dominated by electron-electron scattering[16], (ii) an intermediate temperature regime ($2.8 < \alpha < 3.2$) as a cross-over to a (iii) high temperature regime ($\alpha \approx 2$), where charge carrier scattering is distinguished from normal metal behavior by effects of correlation, the three-fold Fermi surface and intrinsic anisotropic scattering in *k*-space[16].

A first important outcome is that the high-quality thin epitaxial films of CaVO$_3$ on LaAlO$_3$ compare well with the best single crystals[1,2]. However, the range of lowest temperatures in the regime (i) which shows a constant residual resistivity is reduced by almost a factor 10 for the epitaxial films (4 K) as compared to that of single crystals[1,2] (30 K). This indicates that finite size effects may play a role in enhancing electron-phonon scattering as the effective mean free path exceeds the film thickness (Fig. 1(d)).

Similarly, the magnetoresistance results compare well with those reported before[1,2,16]. A positive magnetoresistance is indicative for stoichiometric $CaVO_3$ with standard magnetic-field induced carrier localization. In thin films, finite size effects and substrate-induced tensile strain (0.47%) may be the origin for an *enhanced* magnetoresistance (32% at 5 T, 4 K) as compared to single crystals[1] (2% at 5 T, 2 K).

From the power law behavior (exponent $\beta$) of the field dependence of the magnetoresistance at different temperatures (Fig. 3) we identify three temperature regimes: (I) ($\beta \approx 1$) at low temperatures up to 20 K, where a non-saturating LMR dominates, corresponding to the temperature regime (i) of the Fermi liquid behavior, and the resistivity is dominated by the second electron channel identified by the multi-band fit (Fig. 4); (II) ($\beta \approx 1.33$) in an intermediate temperature regime (20 – 60 K) and (III) ($\beta \approx 1.65$) in the high-temperature regime, with a dominant $B^2$-term above 60 K. This kind of analysis is not yet widespread for $CaVO_3$; however, it reveals clearly a strongly enhanced $MR$ at low temperatures distinguishing the ultraclean and the disordered limit. In the ultraclean limit decreased defects densities may enhance transport in open orbits on the Fermi surface. Additionally, in epitaxial films the electronic bandwidth and local site environment may be sensitive to charge redistribution, confinement and epitaxial strain inducing changes in the V-O bond lengths and V-O overlap/hybridization[11].

The ultimate proof that we have approached the ultraclean limit, is given by the Shubnikov-de Haas oscillations with a frequency of 52 T (Fig. 6(b)) at high magnetic fields, which we attribute to the hole channel. The occurrence of quantum oscillations confirms a crystalline quality equivalent to single crystals[2,27]. For $CaVO_3$ single crystals[27] a de-Haas-van-Alphen frequency of 5000 T is reported; however, this frequency cannot be observed within the magnetic field range up to 12 T in this study. Further, off-stoichiometric $CaVO_{3-y}$ single crystals[2] show a SdH oscillation frequency of 300 T: here, the difference to our films may be accounted for by oxygen vacancies.

Furthermore, the SdH oscillations distinguish the epitaxial CaVO$_3$ films from epitaxial SrVO$_3$ films[16], for which no quantum oscillations were reported, despite the fact that for ultraclean SrVO$_3$ the carrier concentration of the minority hole is large enough to cause a significant change in the Hall derivative[16]. Next to strain, the main difference lies in the exact perovskite structure (cubic vs. pseudo-cubic) of the two correlated metals. However, resolving this issue for SrVO$_3$ will require further investigations.

In the following, we discuss the non-saturating linear magnetoresistance (LMR). It is a peculiar feature of the ultraclean limit for all systems considered here[1,2,16]. It appears strongly reduced for the disordered limit and does not occur for polycrystalline films[34]. Classically, open orbits are the main cause of the non-saturating $MR$[43] which has been previously proposed[1] as mechanism for CaVO$_3$. In the multi-band model[10,30] the magnetoresistance scales in general with $B^2$ and can have a nearly linear $B$-dependence only in a small magnetic-field range[44]. However, the LMR is particularly observed for $\mu B > 1$. A LMR may originate from different effects. First, averaging of anisotropic mobility in polycrystalline materials[45,46], however, polycrystallinity can be ruled out in this study and furthermore, for polycrystalline CaVO$_3$ films no LMR was reported[34]. Second, the linear *quantum* magnetoresistance[47] is rather unlikely, because it requires a majority charge carrier density of the order of $10^{18}$ cm$^{-3}$. Third, multi-band transport in combination with anisotropies, such as in SrMnBi$_2$ with Dirac and parabolic band states, shows non-saturating LMR[48], in which the LMR may be attributed to anisotropic scattering between the different bands. Fourth, anisotropies in *k*-space: orbits around sharp edges on the Fermi surface (e.g. square-like Fermi surfaces) appear as increased elastic scattering, while rounded corners produce a low-field quadratic *MR* dependence[45]. This is consistent for the observed LMR in the low-field regime ($\mu B < 1$) of CaVO$_3$ and SrVO$_3$[16,31], where three Fermi surfaces exist, and the outer Fermi surface is a cuboid with cylindrical openings on its side surfaces. The two inner Fermi surfaces are spherical and cubic, respectively, with the cubic surface potentially causing LMR due to its sharp edges. However, in this model the LMR does not persist for $\mu B > 1$ and saturates above $\mu B \approx 0.6$. Alternatively, impeded orbital motion[49] has been proposed to account for non-saturating LMR in the high-field regime, whereby the scattering time of the carriers undergoes a drastic change periodically. Regarding our results, this model appears to provide a possible picture consistent with multi-carrier transport in the presence of sharp edges in the three-fold Fermi surface of CaVO$_3$.

In order to gain further insights, angle-dependent magnetoresistance measurement may be used to explore scattering anisotropies, and magnetic field investigations above 12 T may reveal quantum oscillations from the electron channels, and show whether the magnetoresistance saturates.

## IV. CONCLUSION

In summary, we demonstrate Shubnikov-de Haas oscillations and an enhanced non-saturating linear magnetoresistance in thin coherently strained epitaxial CaVO$_3$ films. Our transport characterization of a series of thin films with increasing $RRR$ (from 2 to 90) as a measure of increasing epitaxial quality, reveals that for highest film quality (ultraclean limit) electron-like carriers have a mobility of up to 6600 cm$^2$ V$^{-1}$s$^{-1}$ at 4.2 K, not observed for low $RRR$. Quantum oscillations are also intrinsic to CaVO$_3$ single crystals; however, they do not appear in comparable high-quality epitaxial SrVO$_3$ films ($RRR = 195$) prepared by the same method. This calls for detailed investigations regarding origins from the orthorhombic versus the cubic perovskite structure. A non-saturating linear magnetoresistance appears to originate from *intrinsic* anisotropies and a multi-band transport, as it similarly occurs in $CaVO_3$ single crystals and comparable $SrVO_3$ thin epitaxial films. Our findings provide experimental evidence that high-quality epitaxial films of $CaVO_3$, despite coherently substrate-induced small strain and increased surface scattering, display the same features as the bulk correlated metal. Transport phenomena show the delicate interplay of multiple carriers with correlations stemming from a non-spherical nested Fermi surface of a perovskite structure with orthorhombic distortion. These results thus provide fundamental insights, potentially applicable to ultra-thin film transparent electronics.

**Author's contributions**

M. M., M. E., O. C., and S. F. F. conceived and planned the experiments. T. K. and R. E.-H. performed the film growth. M. M. and M. E. performed the transport experiments. All authors interpreted the data. M. M., O. C., R. E.-H., and S. F. F. wrote the manuscript with the input from all authors.

**Acknowledgements**

M. M., M. E., R. E.-H. and S. F. F. acknowledge the support from the Leibniz-Science Campus GraFOx-II, funded by the Leibniz Association (W40/2019) and the Humboldt-Universität zu Berlin. T. K. and R. E.-H. acknowledge the support from the National Science Foundation through DMREF Grant No. DMR-1629477 for the film growth, and structural characterization. The magnetotransport data up to 12 T were taken in the Joined Lab for Advanced Magnetotransport Adlershof.

## Experimental details

The CaVO$_3$ thin films were grown on 10 mm × 10 mm LaAlO$_3$ (100) substrates ($a$ = 3.792 Å) in a DCA M600 hybrid molecular beam epitaxy system. All details about growth and structure can be found in Ref. 21. The calcium flux of $2.5 \cdot 10^{13}$ atoms cm$^{-2}$ s$^{-1}$ was supplied via a solid-state effusion cell, while the metalorganic precursor vanadium oxitriisopropoxide (VTIP) was employed as vanadium source. Thin CaVO$_3$ films were grown by co-supplying Ca and VTIP using a range of different VTIP supply fluxes, quantified as VTIP gas inlet pressure $P_{\text{VTIP}}$. Film thickness of 38 ± 2 nm and out-of-plane lattice parameter were extracted from the high resolution XRD scans[21]. Reciprocal space maps obtained from X-ray diffraction indicate that the films are coherently strained on the substrate, exhibiting a tensile strain[21] of 0.47 %. For electrical characterization the films were diced into 5 mm × 5 mm and 1.25 mm × 5 mm pieces, the first of which was used for electrical measurements in Van der Pauw geometry and the second for measurements in Hall geometry. Wire wedge-bonding was used for ohmic contacts to the films with an Al/Si (99%/1%) wire. For the measurements a Keithley 6221/2182A Complete Delta Mode system was used, combined with a Keithley 7001 switch matrix. The resistivity and Hall measurements in Van der Pauw geometry were performed in a KONTI IT Cryostat with a magnetic field of up to 700 mT. The Hall measurements were extended to 6 T using a Quantum Design Physical Property Measurement System (PPMS), with the Model 7100 AC Transport Controller for transport measurements in 5-wire Hall measurement configuration[22]. For the evaluation of the magnetoresistance data the antisymmetric part of the Hall resistance has been subtracted. In both cryogenic systems the ambient conditions were helium atmosphere at standard pressure. For the 12 T measurements an Oxford Instruments TeslatronPT system was used with a KelvinoxJT insert. The measurements were performed in Van der Pauw geometry and the films were kept in vacuum.

For resistivity and Hall analysis the resistance $R$ was measured by ramping a current $I$ from zero to its positive maximum, then to its negative maximum and back to zero. Meanwhile, the voltage was measured and the resistance was determined from the slope $R = \frac{\partial U}{\partial I}$. A sweep consisted of at least 100 points and the uncertainty for the slope of the linear model gives the uncertainty for the resistance. With this method, the uncertainty for all resistance measurements is below 0.1 %.

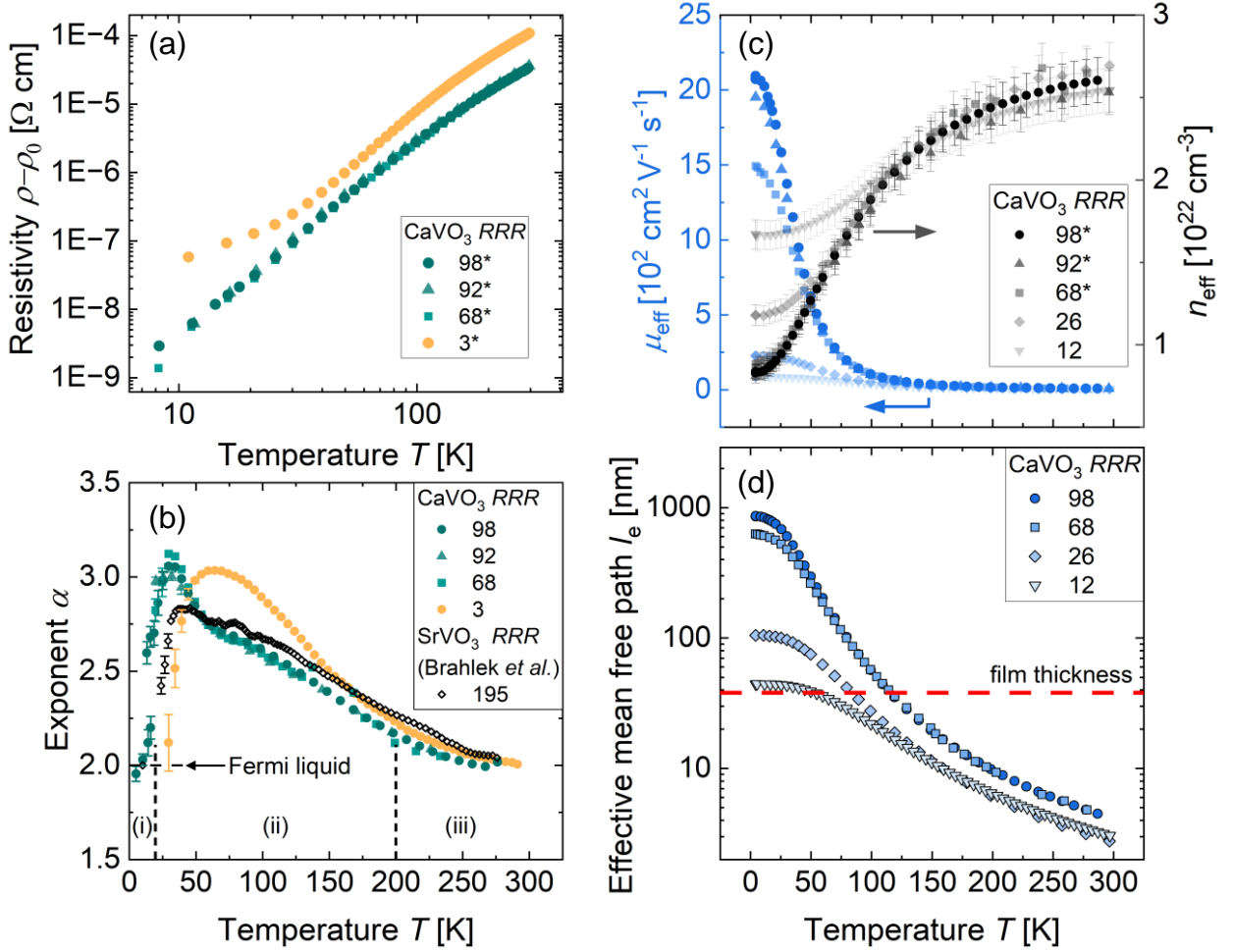

Figure 1 (a) Temperature dependent resistivity $\rho(T)$ for films with different $RRR$. $\rho_0$ is the residual resistivity. Data marked with ($\star$) were previously published in Ref. 21. (b) Exponent $\alpha$ extracted from the resistivity fit $\rho(T) = \rho_0 + A \cdot T^\alpha$ for four different CaVO$_3$ films, and an ultraclean SrVO$_3$ film taken from Ref. 16. (c) Low-field inverse Hall coefficient $n_{\text{eff}} = (-eR_{\text{H}})^{-1}$ and effective mobility $\mu_{\text{eff}}$ evaluated within the one-band model. For higher $RRR$ values, the decrease of the inverse Hall coefficient and the increase in effective mobility is largest. (d) Calculated *effective* mean free path $l_e$ (Drude picture). For films with highest $RRR$, $l_e$ significantly exceeds the film thickness $d$ (red dashed horizontal line).

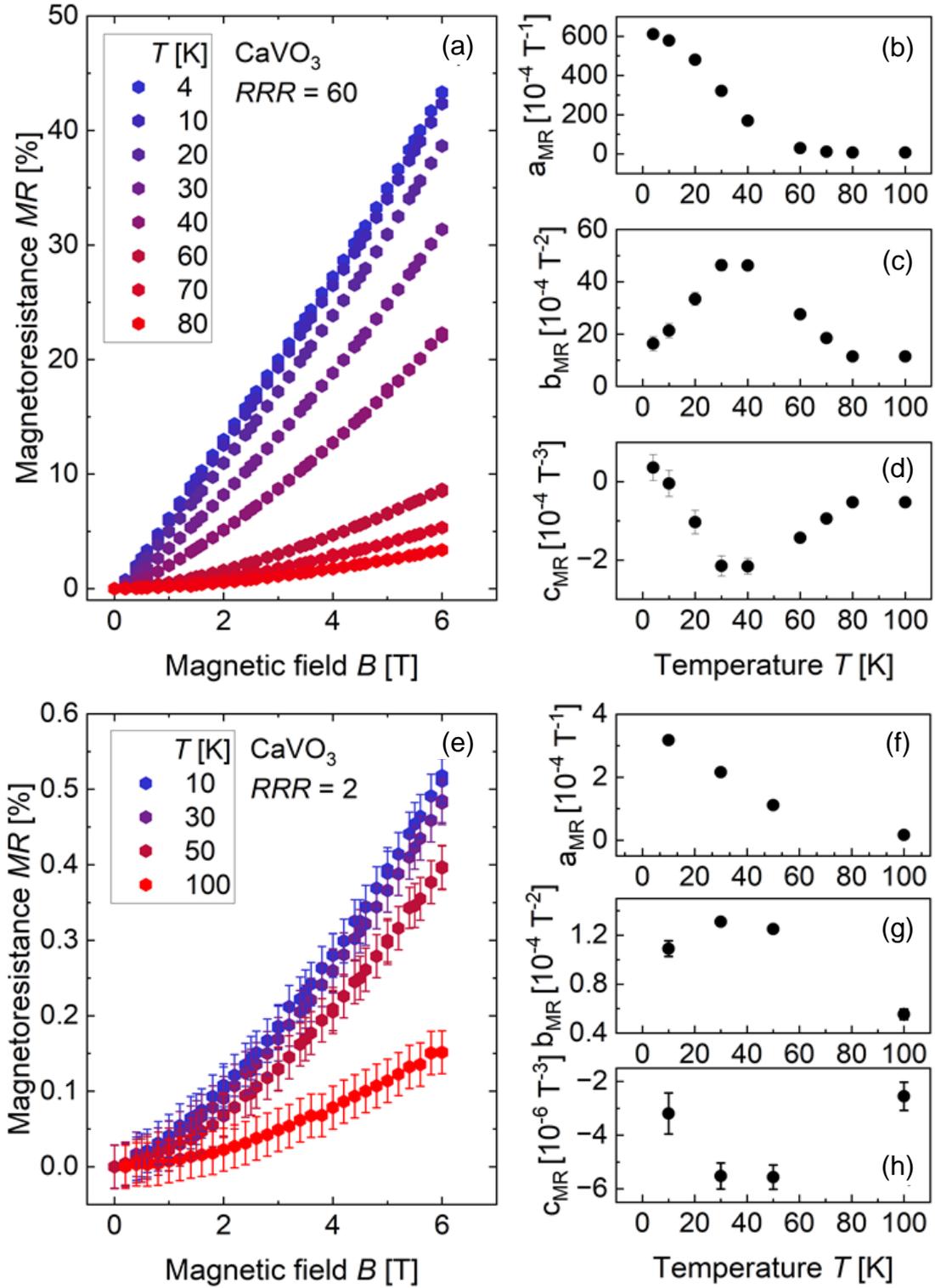

Figure 2 Magnetoresistance of $CaVO_3$ epitaxial thin films with (a) $RRR = 60$ and (e) $RRR = 2$ for different temperatures. The corresponding fit parameters for $MR = a_{MR}|B| + b_{MR} \cdot B^2 + c_{MR} \cdot |B^3|$ are shown in (b)-(d) and (f)-(h), respectively.

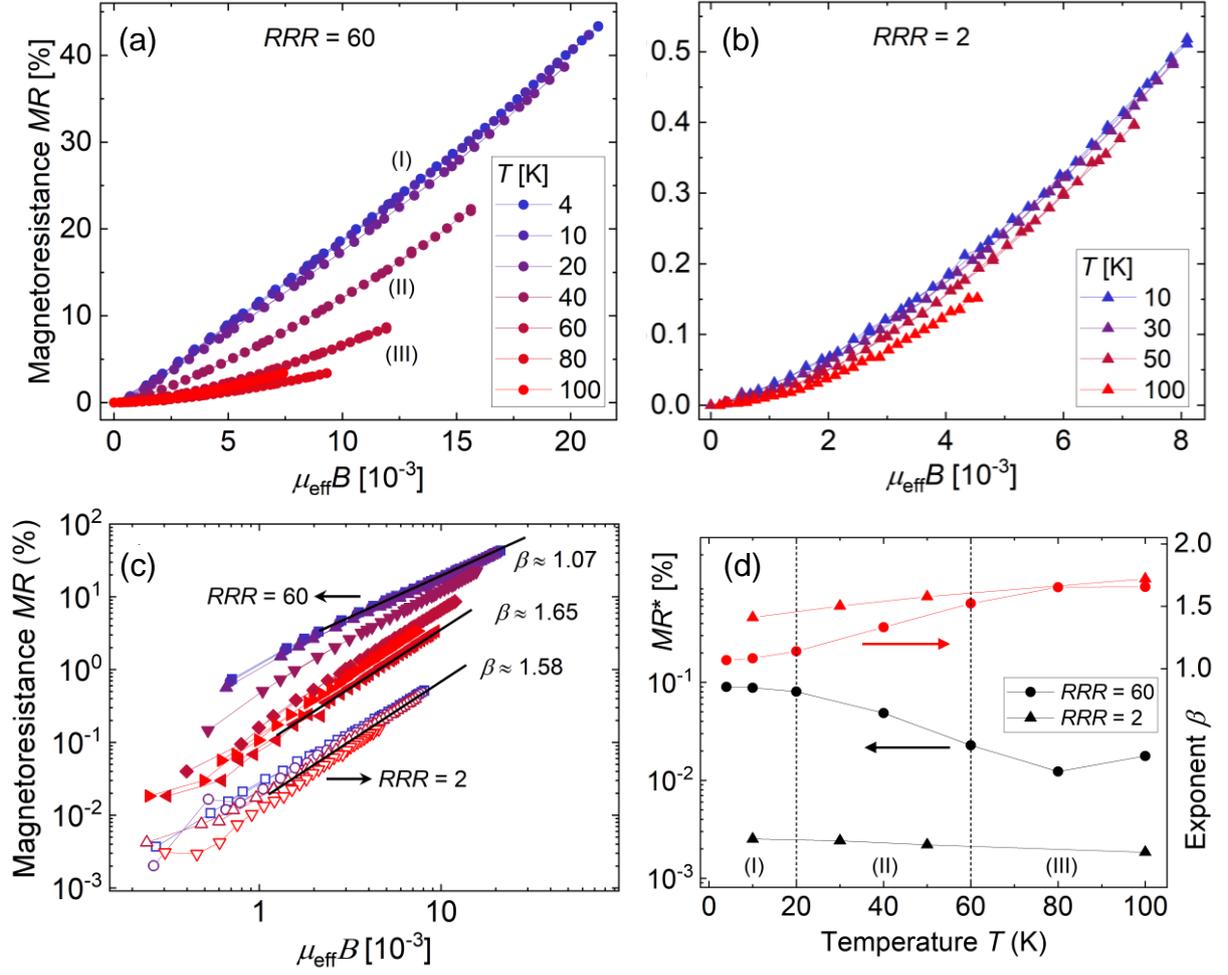

Figure 3 Magnetoresistance as a function of normalized magnetic field $\mu_{\text{eff}}B$ for CaVO$_3$ epitaxial thin films with (a) $RRR = 60$ and (b) $RRR = 2$, for different temperatures. (c) Data from (a) and (b) in a log-log plot to emphasize the power-law dependence on the magnetic field: $MR \propto B^\beta$. The solid lines are guides to the eye for the corresponding exponent. (d) Magnetoresistance $MR^*$ at $\mu_{\text{eff}}B \approx 5 \cdot 10^{-3}$ and exponent $\beta$ as a function of temperature.

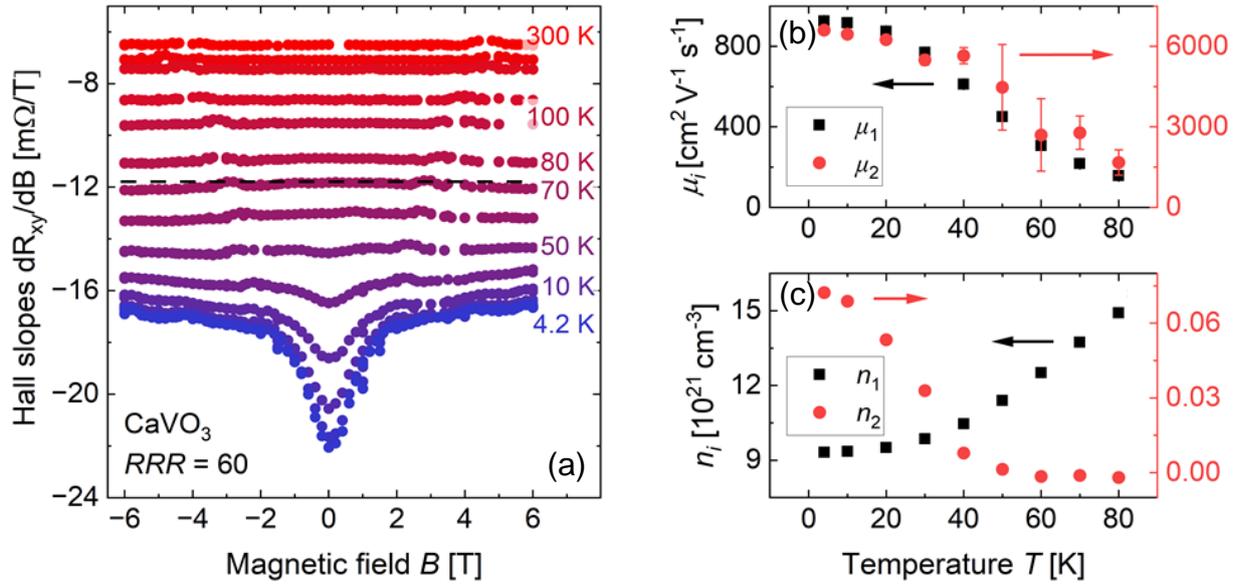

Figure 4 CaVO$_3$ epitaxial thin films: (a) Derivative of the Hall resistance $\frac{\partial R_{xy}}{\partial B}$ as a function of magnetic field for $RRR = 60$. (b) and (c) Results of the two-band fits of the data in (a) for the temperature-dependent mobility and carrier concentration, respectively, of two electron channels.

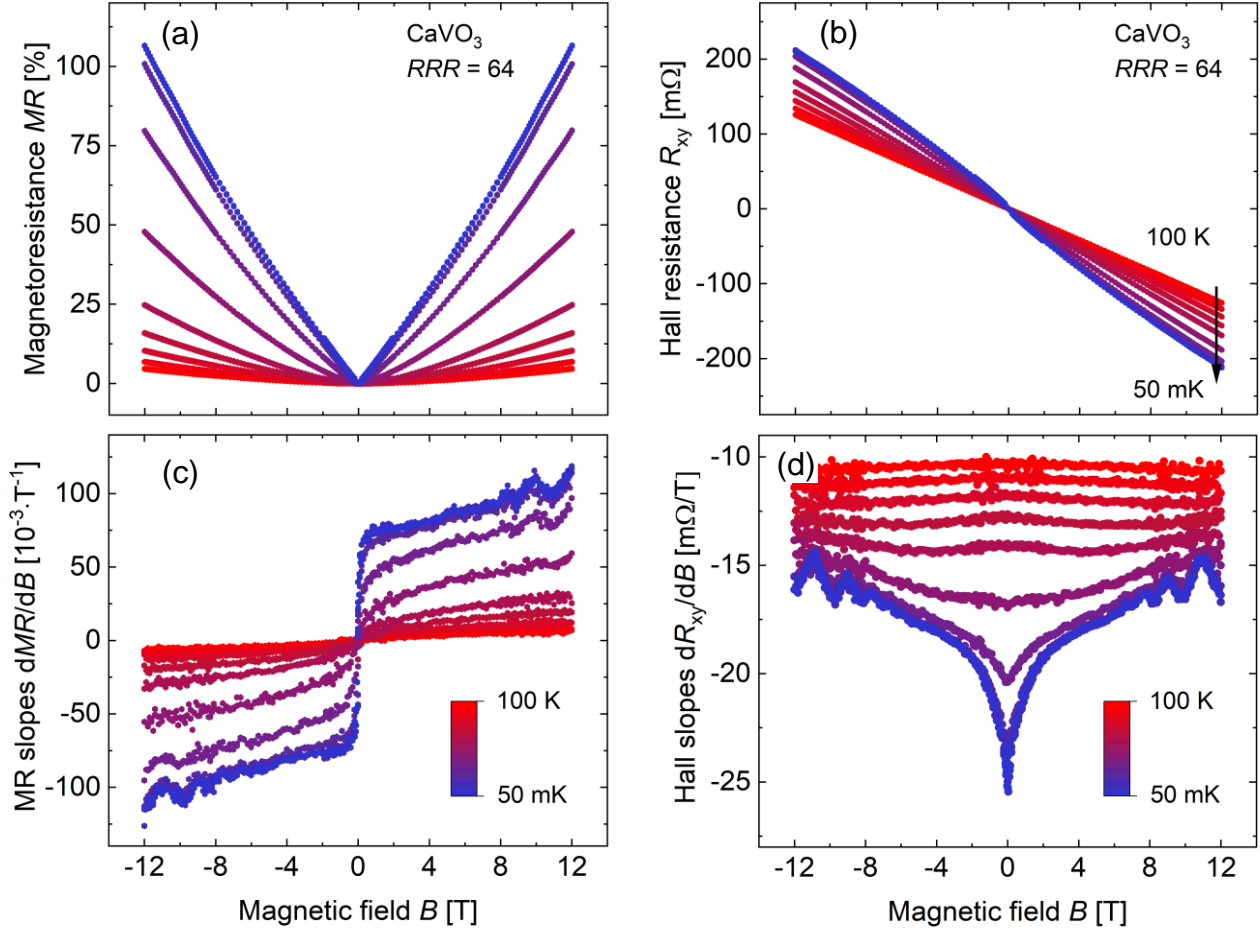

Figure 5 (a) The magnetoresistance for a CaVO$_3$ epitaxial thin film and (c) the mean derivative of the magnetoresistance in magnetic fields up to 12 T. (b) Hall resistances at different temperatures for a CaVO$_3$ epitaxial thin film and (d) the mean derivative for every temperature in magnetic fields up to 12 T. The temperatures are 100, 90, 80, 70, 60, 45, 30, 15, 0.8 and 0.05 K, respectively.

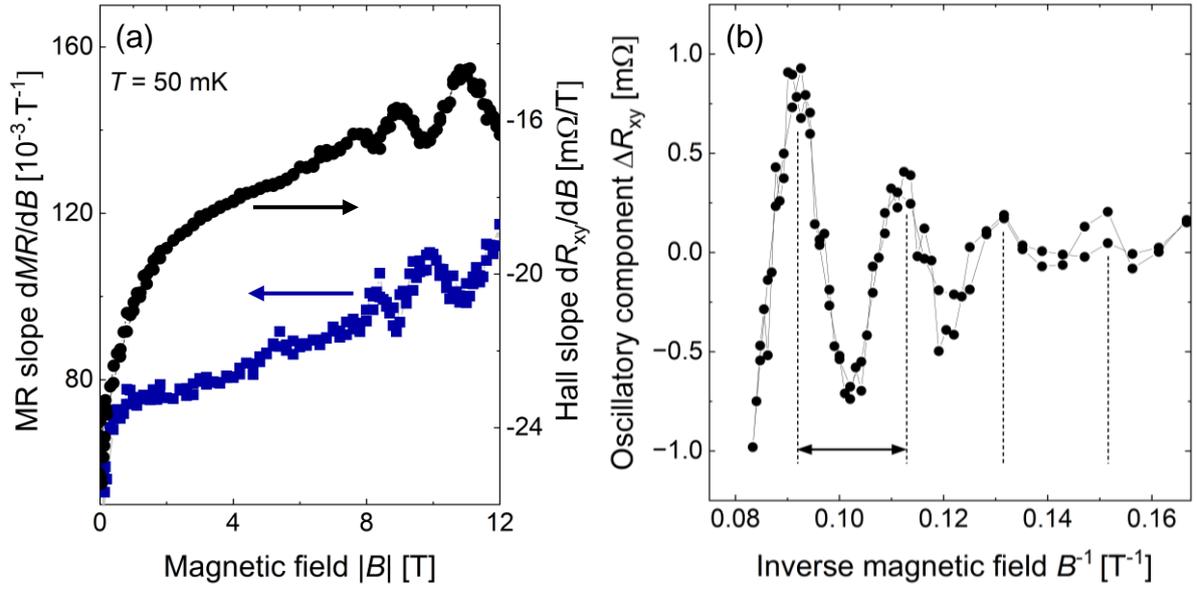

Figure 6 CaVO$_3$ epitaxial thin film with $RRR = 64$: (a) Magnetoresistance and Hall slope as a function of the absolute value of the magnetic field at $T = 50$ mK, showing oscillations for $B > 7$ T. (b) Oscillatory component of the Hall resistance as a function of inverse magnetic field $B^{-1}$ at $T = 50$ mK. The Shubnikov-de Haas period is estimated to be $\Delta\left(\frac{1}{B}\right) = 0{,}019$ T, corresponding to a frequency of $B_{\text{SdH}} \approx 52$ T.